\documentclass{moriond_mod}

\usepackage{amsmath}
\usepackage{amssymb}

\def\be{\begin{equation}}
\def\ee{\end{equation}}
\def\bea{\begin{eqnarray}}
\def\eea{\end{eqnarray}}

\newcommand{\crn}{\nonumber \\}

\newcommand{\eq}[1]{Eq.~(\ref{#1})}
\newcommand{\bib}[1]{Ref.~\cite{#1}}

\newcommand{\fig}[1]{Fig.~\ref{#1}}
\newcommand{\tab}[1]{Table~\ref{#1}}

\newcommand{\sect}[1]{Section~\ref{#1}}

\newcommand{\Photo}{}

\begin{document}
\rightline{IFIRSE-TH-2019-2}
\vspace*{4cm}
\title{POLARISATION AT NLO IN \boldmath $WZ$ PRODUCTION AT THE LHC}

\author{ \underline{JULIEN BAGLIO}\,\footnote{Speaker} $^{1,2}$ AND LE DUC NINH $^{2}$}

\address{$^1$ Institut f\"{u}r Theoretische Physik, Eberhard Karls Universit\"{a}t
T\"{u}bingen,\\ Auf der Morgenstelle 14, D-72076 T\"{u}bingen, Germany}
\address{$^2$ Institute for Interdisciplinary Research in Science and Education,\\  
ICISE, 590000 Quy Nhon, Vietnam}

\maketitle
\abstracts{
The pair production of a $W$ and a $Z$ boson at the LHC is an
important process to study the triple-gauge boson couplings as well as
to probe new physics that could arise in the gauge sector. In particular
the leptonic channel $p p \to W^\pm Z\to 3\ell +\nu + X$ is considered by
ATLAS and CMS collaborations. Polarisation observables can help
pinning down new physics and give information on the spin of the
gauge bosons. Measuring them requires high statistics as well as
precise theoretical predictions. We define in this contribution
fiducial polarisation observables for the $W$ and $Z$ bosons and we
present theoretical predictions in the Standard Model at
next-to-leading order (NLO) including QCD as well as NLO electroweak
corrections, the latter in the double-pole approximation. We also show
that this approximation works remarkably well for $W^\pm Z$ production
at the LHC by comparing to the full results.}

\section{Introduction}
\label{sec:intro}

The production of a $WZ$ pair at the LHC, in the fully leptonic
channel $pp \to W^\pm Z \to 3\ell + \nu + X$, is an important part of
the Large Hadron Collider (LHC) programme to test the electroweak (EW)
sector of the Standard Model (SM). In particular, this process is
sensitive already at leading order (LO) to the triple-gauge-boson
couplings that can be modified by new physics. High statistics than
can be collected at the LHC allows for precise measurements even in
complex observables such as kinematical distributions and polarisation
observables. This also requires precise predictions from the theory
side. The next-to-leading order (NLO) QCD corrections were calculated
for the first time in the
nineties~\cite{Ohnemus:1991gb,Frixione:1992pj}, while the
next-to-next-to-leading order QCD corrections became available
recently, including off-shell
effects~\cite{Grazzini:2016swo,Grazzini:2017ckn}. The NLO EW
corrections to the on-shell $WZ$ production were first calculated in
Refs.~\cite{Bierweiler:2013dja,Baglio:2013toa} and the full NLO EW
corrections to the process $p p\to 3\ell + \nu +X$ have been recently
computed providing results for various kinematical
distributions~\cite{Biedermann:2017oae}.

Even if the initial proton beams at the LHC are unpolarised, the
fundamental asymmetry in the $Z$ and $W$ boson couplings to
left- and right-handed quarks means that the gauge bosons are produced
in polarised states. This polarisation is reflected in angular
asymmetries in the final-state-lepton distributions. First
measurements of these polarisation observables at the 13 TeV LHC have
been presented by ATLAS in 2019~\cite{Aaboud:2019gxl}.

We will present in the following polarisation observables built out
from the polar-azimuthal angular distributions, that can be defined in
the fiducial phase-space and that reflect the underlying spin
structure of the $W$ and $Z$ bosons. These fiducial polarisation
observables allow for direct comparisons with the experiment, without
using template fitting. They will be calculated at NLO in QCD and also
at NLO EW in the double-pole approximation. We will also show
that the DPA for the EW corrections works remarkably well in this
process and is a good approximation to perform the NLO calculation of
polarisation. All the details of the analysis can be found in the
original publication~\cite{Baglio:2018rcu}.

\section{EW corrections in the double-pole approximation}
\label{sec:calcsetup}

We calculate the cross section and the kinematical distributions, in
particular the fiducial distribution $d\sigma/(\sigma d\!\cos\theta
d\phi)$, at LO and NLO QCD using the full matrix elements. We use the
{\tt VBFNLO} program~\cite{Arnold:2008rz,Baglio:2014uba} to generate
the LO and NLO QCD results. The NLO EW corrections are obtained using
the double pole approximation (DPA) in which the on-shell $W^\pm Z$
production and the on-shell decays $W\to e \nu_e$, $Z\to \mu^+ \mu^-$ 
are combined. At LO, the partonic amplitude in the DPA for the process
$a b\to e \nu_e\mu^+\mu^-$ reads
\begin{align}
\mathcal{A}_{\rm LO,\ DPA}^{a b\to W Z\to e\nu_e \mu^+\mu^-} =
  \sum_{\lambda_1,\lambda_2}^{} \frac{\mathcal{A}_{\rm LO}^{a b\to W
  Z}\ \mathcal{A}_{\rm LO}^{W\to e\nu_e}\ \mathcal{A}_{\rm
  LO}^{Z\to\mu^+\mu^-}}{(q_{W^*}^2-M_W^2+i M_W^{}
  \Gamma_W^{})(q_{Z^*}^2-M_Z^2+i M_Z^{}\Gamma_Z^{})},
\end{align}
where $q_{W^*}$ and $q_{Z^*}$ are the intermediate off-shell $W$ and
$Z$ momenta, $\lambda_1$ and $\lambda_2$ are the helicities of
the intermediate on-shell $W$ and $Z$ bosons, respectively.
At NLO EW we take over the corrections to the production
part from \bib{Baglio:2013toa} and we calculate the NLO EW corrections
to the decay parts. When compared to the full calculation of the EW
corrections~\cite{Biedermann:2017oae}, the virtual and real corrections
of the production and of the decays parts are included, in particular
the quark-photon induced $q\gamma\to W^\pm Z q'\to e\nu_e\mu^+\mu^-
q'$ channel, while the non-factorisable contributions as well as the
off-shell effects are neglected. A great advantage of the DPA is also 
the lesser time spent in the calculation, as we avoid calculating
six-point integrals.

\begin{figure}[ht!]
  \centering
  \begin{tabular}{cc}
    \includegraphics[width=0.48\textwidth]{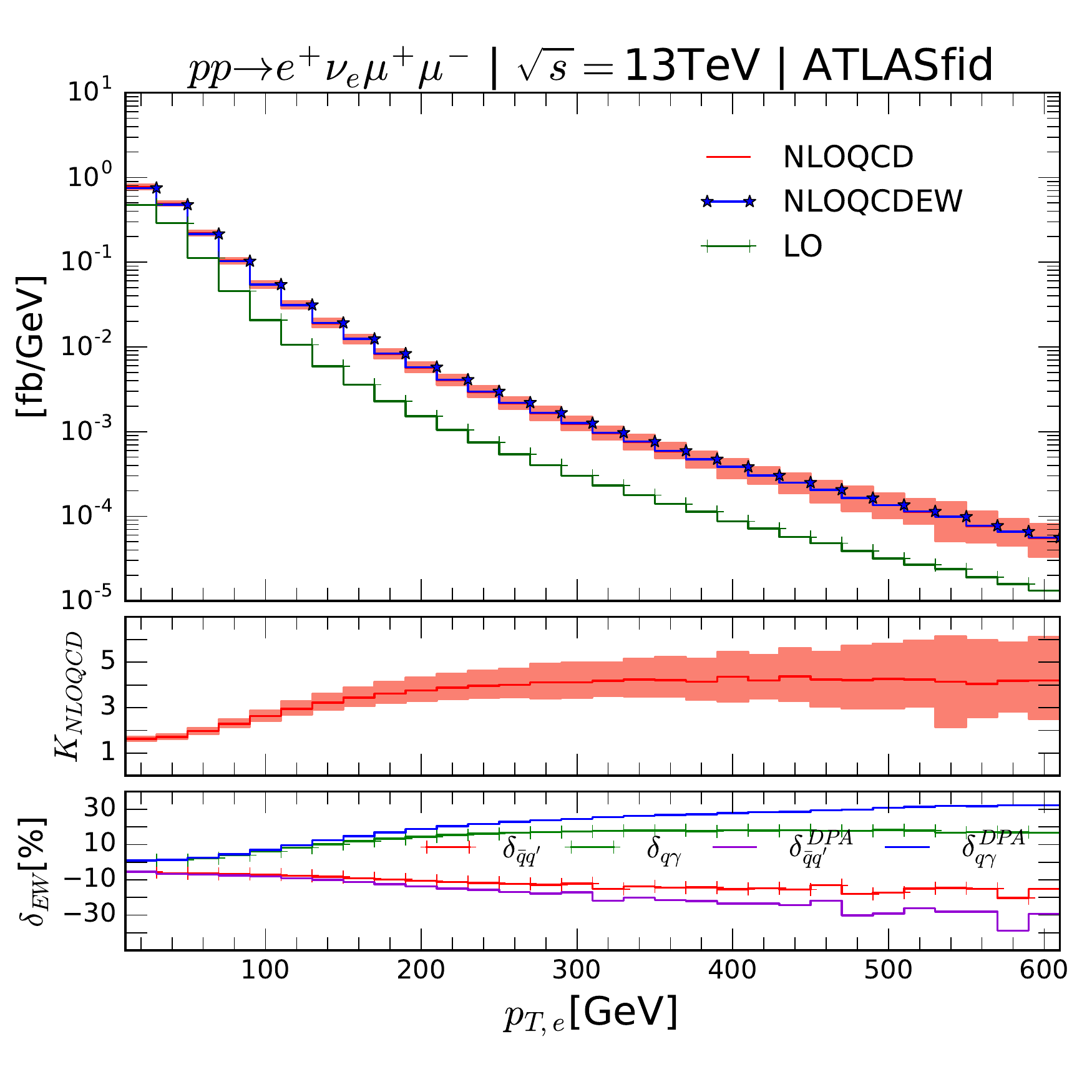}
    &
      \includegraphics[width=0.40\textwidth]{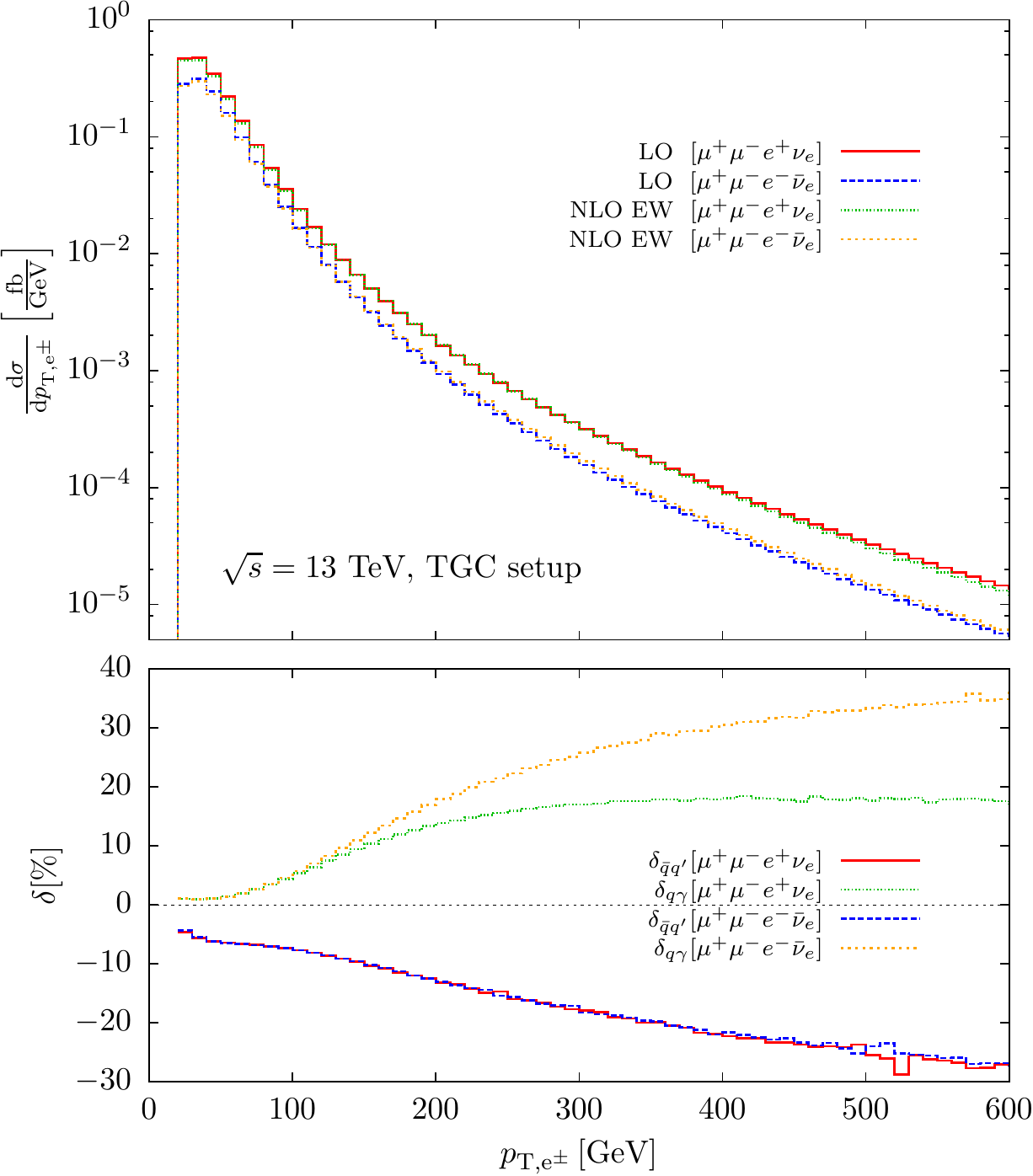}
  \end{tabular}
  \caption{Transverse momentum distributions of the positron in the
    process $p p\to e^+\nu_e\mu^+\mu^- + X$ at the 13 TeV LHC using ATLAS
    fiducial cuts. Left: Our calculation~\protect\cite{Baglio:2018rcu}
    including full NLO QCD corrections only (in red) and their
    combination with EW corrections in the DPA (in blue) in the main
    panel, as well as the percent corrections of the EW corrections in
    the lower sub-panel. Right: The full NLO EW calculation taken from
    \protect\bib{Biedermann:2017oae}.}
  \label{fig:pTeATLAS}
\end{figure}

We display in Fig.~\ref{fig:pTeATLAS} a comparison of our results in
the DPA with the calculation of the full EW corrections, at the 13 TeV
LHC using ATLAS fiducial cuts~\cite{Aaboud:2016yus}, using the $p_T$ 
distribution of the positron in the $W^+Z$ channel as an illustrative
example. The agreement is very good, signalling that the DPA is an
excellent approximation that is suitable for the calculation of the
polarisation observables.

\section{Fiducial polarisation observables}
\label{sec:fiducial_polarization}

At LO in the DPA it is possible to parameterise the differential cross
section as
\begin{align}
\frac{1}{\sigma_{\rm LO}^{\rm DPA} }\frac{d\sigma_{\rm LO}^{\rm DPA}}{d\!\cos\theta d\phi} &= \frac{3}{16\pi}
\Big[ 
(1+\cos^2\theta) + A_0 \frac{1}{2}(1-3\cos^2\theta)
+ A_1 \sin(2\theta)\cos\phi  \crn
& + A_2 \frac{1}{2} \sin^2\theta \cos(2\phi)
+ A_3 \sin\theta\cos\phi + A_4 \cos\theta \crn
& + A_5 \sin^2\theta \sin(2\phi) 
+ A_6 \sin(2\theta) \sin\phi + A_7 \sin\theta \sin\phi
\Big],\label{eq:definition_Ai}
\end{align}
where $\theta$ and $\phi$ are the angular variables of one of the
final-state lepton in the gauge-boson rest frame, and $A_i$ are called
the angular coefficients. To study the $W^\pm$ boson the electron (or
positron) is taken, to study the $Z$ boson we chose the $\mu^-$
lepton.

A polarised on-shell massive gauge boson can be described by a
$3\times 3$ spin-density matrix $\rho_{ij}$. This matrix is Hermitian
and satisfies the normalisation condition of $\text{Tr}(\rho)=1$,
hence it can be parameterised by $8$ real parameters that characterise
the spin nature of the massive gauge boson. At LO in the DPA, the link
between the angular coefficients and the spin-density matrix is given
by~\cite{Aguilar-Saavedra:2015yza,Aguilar-Saavedra:2017zkn}
\begin{align}
A_0 &= 2 \rho_{00},\; A_1 = \frac{1}{\sqrt{2}}(\rho_{+0}-\rho_{-0}+\rho_{0+}-\rho_{0-}),\crn
A_2 &= 2(\rho_{+-} + \rho_{-+}),\; A_3 = \sqrt{2}b(\rho_{+0} + \rho_{-0} + \rho_{0+} + \rho_{0-}),\crn
A_4 &= 2b(\rho_{++} - \rho_{--}),\; A_5 = \frac{1}{i}(\rho_{-+} - \rho_{+-}),\crn
A_6 &= -\frac{1}{i\sqrt{2}}(\rho_{+0} + \rho_{-0} - \rho_{0+} - \rho_{0-}),\; 
A_7 = \frac{\sqrt{2}b}{i}(\rho_{0+} - \rho_{0-} - \rho_{+0} + \rho_{-0}),
\label{eq:Ai_spin_matrix}
\end{align}  
with $b=1$ for the $W^\pm$ bosons and $b=-c$ for the $Z$ boson, and
\bea
c = \frac{g_L^2 - g_R^2}{g_L^2 + g_R^2} =
\frac{1-4s^2_W}{1-4s^2_W+8s^4_W},\quad s^2_W = 1 - \frac{M_W^2}{M_Z^2}.
\label{eq:def_c}
\eea
When looking at \eq{eq:Ai_spin_matrix} it is expected that the
coefficients $A_{5,6,7}$ are small as they are related to the
imaginary part of the spin-density matrix. Furthermore, $A_3^Z$ and
$A_4^Z$ are proportional to the coefficient $c$ and are thus directly
related to the left-right asymmetry in the decay $Z^*\to
\mu^+\mu^-$. It is possible to relate the angular coefficients to the
angular projections of the two-dimensional distribution
$d\sigma_{\rm LO}^{\rm DPA}/d\!\cos\theta
d\phi$~\cite{Bern:2011ie,Stirling:2012zt},
\begin{align}
A_0 &= 4 - \langle 10\cos^2\theta\rangle, \; A_1 = \langle 5\sin 2\theta\cos\phi \rangle, 
\; A_2 = \langle 10\sin^2\theta \cos 2\phi \rangle, \; A_3 = \langle 4 \sin\theta\cos\phi \rangle,\crn 
\; A_4 &= \langle 4\cos\theta\rangle, \; A_5 = \langle 5\sin^2\theta\sin 2\phi \rangle, 
\; A_6 = \langle 5\sin 2\theta \sin\phi \rangle, \; A_7 = \langle 4\sin\theta \sin\phi \rangle,
\label{eq:Ai_proj}
\end{align}
with the following definition for angular projections
\begin{align}
\langle g(\theta,\phi) \rangle = \int_{-1}^{1} d\!\cos\theta \int_{0}^{2\pi} d\,\!\phi
g(\theta,\phi)\frac{1}{\sigma}\frac{d\sigma}{d\!\cos\theta d\phi}.
\label{eq:defs_proj} 
\end{align}

If off-shell and radiation effects as well as cuts on the individual
leptons are considered, as in the full NLO calculation in the fiducial
region, the parameterisation in \eq{eq:definition_Ai} is not valid
anymore: It only contains information on the $Z$ and $W$
bosons and the eight coefficients $A_i$ cannot reconstruct the full
differential cross section anymore. The problem gets worse when
considering the same-flavour final states of $\ell \nu_\ell \ell^+
\ell^-$, where interference effects occur. Nevertheless it is always
possible to use the two-dimensional distribution and the definitions
given in \eq{eq:Ai_proj}. In this equation it is even possible to
replace the cross section $\sigma$ by a differential distribution such
as $d\sigma/d p_{T,W/Z}$~\cite{Stirling:2012zt}. In this way we can
define what we call fiducial polarisation observables that are totally
equivalent to the inclusive polarisation observable in the DPA limit
in the inclusive phase-space. The fiducial polarisation observables
are calculated using the full matrix elements including off-shell,
interference, and higher-order effects as well as arbitrary cuts on
the individual final-state leptons, and they do contain the spin
information of the underlying $W$ and $Z$ bosons. The fiducial
polarisation observables are measurable and they do not require a
template fit method, contrary to what is done in
\bib{Aaboud:2019gxl}.

It is also possible to define polarisation fractions as
\begin{align}
f^V_L 
  = -\frac{1}{2} + d\langle\cos\theta\rangle +
    \frac{5}{2}\langle\cos^2\theta\rangle, \;
f^V_R 
  = -\frac{1}{2} - d\langle\cos\theta\rangle +
    \frac{5}{2}\langle\cos^2\theta\rangle, \;
f^V_0 
  = 2 - 5 \langle\cos^2\theta\rangle,
\label{eq:polfrac}
\end{align}
with $d=\mp 1 $, $\theta = \theta_{e}$ for $V=W^\pm$ and $d=1/c$,
$\theta = \theta_{\mu^-}$ for $V=Z$. The fractions fulfil the
relation $f_L+f_R+f_0 = 1$. Their value as well as the value of the
angular coefficients $A_i$ depend on the reference frame and on the
coordinate system. Our study is made in the rest frame of the massive
gauge boson under study, using either the helicity coordinate
system (HE)~\cite{Bern:2011ie} in which the $z$-direction is aligned
with the gauge boson momentum in the laboratory frame, or the
Collins-Soper coordinate system (CS)~\cite{Collins:1977iv} in which the
$z$-direction is the bisection of the momenta of the colliding protons
in the gauge boson rest frame, see \bib{Baglio:2018rcu} for more details.

\section{Results}
\label{sec:results}

We now present the results for the angular coefficients for
the $W^+$ and $Z$ bosons at the 13 TeV LHC with the ATLAS fiducial
cuts~\cite{Aaboud:2016yus}. The study in \bib{Baglio:2018rcu}
contains also results for the polarisation fractions and a study of
the CMS cuts as well as a complete study for the $W^- Z$ process. 

\begin{table}[ht!]
 \renewcommand{\arraystretch}{1.3}
\begin{center}
\setlength\tabcolsep{0.03cm}
\fontsize{8.0}{8.0}
\begin{tabular}{|c|c|c|c|c|c|c|c|c|}\hline
$\text{Method}$  & $A_0$ & $A_1$  & $A_2$ & $A_3$ & $A_4$ & $A_5$ & $A_6$ & $A_7$\\
\hline
{\fontsize{6.0}{6.0}$\text{HE NLOQCD}$} & $1.016(1)^{+3}_{-4}$ & $-0.326(2)^{+2}_{-3}$ & $-1.413(2)^{+10}_{-12}$ & $-0.229(1)^{+2}_{-1}$ & $-0.295(7)^{+11}_{-11}$ & $-0.001(1)^{+0.1}_{-0.2}$ & $-0.0002(6)^{+3}_{-2}$ & $0.003(1)^{+1}_{-0.5}$\\
\hline
{\fontsize{6.0}{6.0}$\text{HE NLOQCDEW}$} & $1.017$ & $-0.326$ & $-1.420$ & $-0.229$ & $-0.287$ & $-0.002$ & $-0.002$ & $0.007$\\
\hline\hline
{\fontsize{6.0}{6.0}$\text{CS NLOQCD}$} & $1.513(3)^{+7}_{-7}$ & $0.192(1)^{+2}_{-2}$ & $-0.918(3)^{+2}_{-2}$ & $0.061(4)^{+4}_{-4}$ & $-0.469(6)^{+10}_{-10}$ & $-0.0001(11)^{+0}_{-3}$ & $0.001(0.5)^{+0.3}_{-0.2}$ & $-0.003(0.4)^{+1}_{-1}$\\
\hline
{\fontsize{6.0}{6.0}$\text{CS NLOQCDEW}$} & $1.518$ & $0.189$ & $-0.921$ & $0.065$ & $-0.463$ & $0.0004$ & $0.003$ & $-0.007$\\
\hline
\end{tabular}
\caption{\small ($W$ boson) Fiducial angular coefficients of the
  $e^+$ distribution for the process $p p\to e^+\nu_e\mu^+\mu^-+X$
  at the 13 TeV LHC with the ATLAS fiducial cuts. Results are
  presented in the helicity (HE) and Collins-Soper (CS) coordinates
  systems, at NLO QCD and NLO QCD+EW. The PDF uncertainties (in
  parenthesis) and the scale uncertainties are also provided on the
  last digit of the central prediction.}
\label{tab:coeff_Ai_W_ATLAS}
\end{center}
\end{table}

\begin{table}[ht!]
 \renewcommand{\arraystretch}{1.3}
\begin{center}
\setlength\tabcolsep{0.03cm}
\fontsize{8.0}{8.0}
\begin{tabular}{|c|c|c|c|c|c|c|c|c|}\hline
$\text{Method}$  & $A_0$ & $A_1$  & $A_2$ & $A_3$ & $A_4$ & $A_5$ & $A_6$ & $A_7$\\
\hline
{\fontsize{6.0}{6.0}$\text{HE NLOQCD}$} & $0.985(2)^{+5}_{-6}$ & $-0.306(1)^{+4}_{-3}$ & $-0.734(1)^{+2}_{-2}$ & $0.031(1)^{+2}_{-2}$ & $0.003(1)^{+1}_{-1}$ & $-0.004(1)^{+0.3}_{-0.4}$ & $-0.004(1)^{+0.3}_{-0.2}$ & $0.003(1)^{+0.2}_{-0}$\\
\hline
{\fontsize{6.0}{6.0}$\text{HE NLOQCDEW}$} & $0.986$ & $-0.308$ & $-0.742$ & $0.023$ & $0.001$ & $-0.004$ & $-0.004$ & $0.003$\\
\hline\hline
{\fontsize{6.0}{6.0}$\text{CS NLOQCD}$} & $1.267(2)^{+4}_{-4}$ & $0.221(1)^{+1}_{-1}$ & $-0.455(2)^{+2}_{-2}$ & $-0.021(1)^{+3}_{-3}$ & $0.023(1)^{+1}_{-1}$ & $0.0004(6)^{+2}_{-2}$ & $0.006(0.5)^{+0.2}_{-0.4}$ & $-0.003(1)^{+0}_{-0.1}$\\
\hline
{\fontsize{6.0}{6.0}$\text{CS NLOQCDEW}$} & $1.273$ & $0.218$ & $-0.457$ & $-0.016$ & $0.016$ & $0.001$ & $0.007$ & $-0.003$\\
\hline
\end{tabular}
\caption{\small Same as in \protect\tab{tab:coeff_Ai_W_ATLAS} but for
  the ($Z$ boson) fiducial angular coefficients of the $\mu^-$
  distribution.}
\label{tab:coeff_Ai_Z_ATLAS}
\end{center}
\end{table}

%old ninh below:

The angular coefficients of the $W^+$ boson $A_i^W$ are presented in
\tab{tab:coeff_Ai_W_ATLAS} and those of the $Z$ boson $A_i^Z$ are in
\tab{tab:coeff_Ai_Z_ATLAS}. The results are given in both the HE and
CS coordinate systems. As expected from the analysis in
\sect{sec:fiducial_polarization} for the inclusive polarisation
observables (see \eq{eq:Ai_spin_matrix}) the coefficients $A_{5,6,7}$
are very small and can be neglected in the SM. However, they could
serve as a probe for new physics as any sizeable non-zero values for
these coefficients constitute a striking signal for physics beyond the
SM. The parton distribution function and the scale uncertainties are
found to be very small in both coordinate systems and for both sets of
angular coefficients, signalling an observable under very good
theoretical control.

The NLO EW corrections are found negligible for all $W$ coefficients
in both coordinate systems, while they are significant for $A_3^Z$ and
$A_4^Z$. A more detailed study of the different DPA contributions to
the EW corrections performed in the original
study~\cite{Baglio:2018rcu} shows that these large EW corrections
originate from the radiative corrections to the $Z\to \mu^- \mu^+$
decay, aligning with the remark in \sect{sec:fiducial_polarization}
about the proportionality of $A_3^Z$ and $A_4^Z$ to the left-right
asymmetry in the $Z\to\mu^+\mu^-$ decay.

\begin{figure}[ht!]
  \centering
  \begin{tabular}{cc}
    \includegraphics[width=0.48\textwidth]{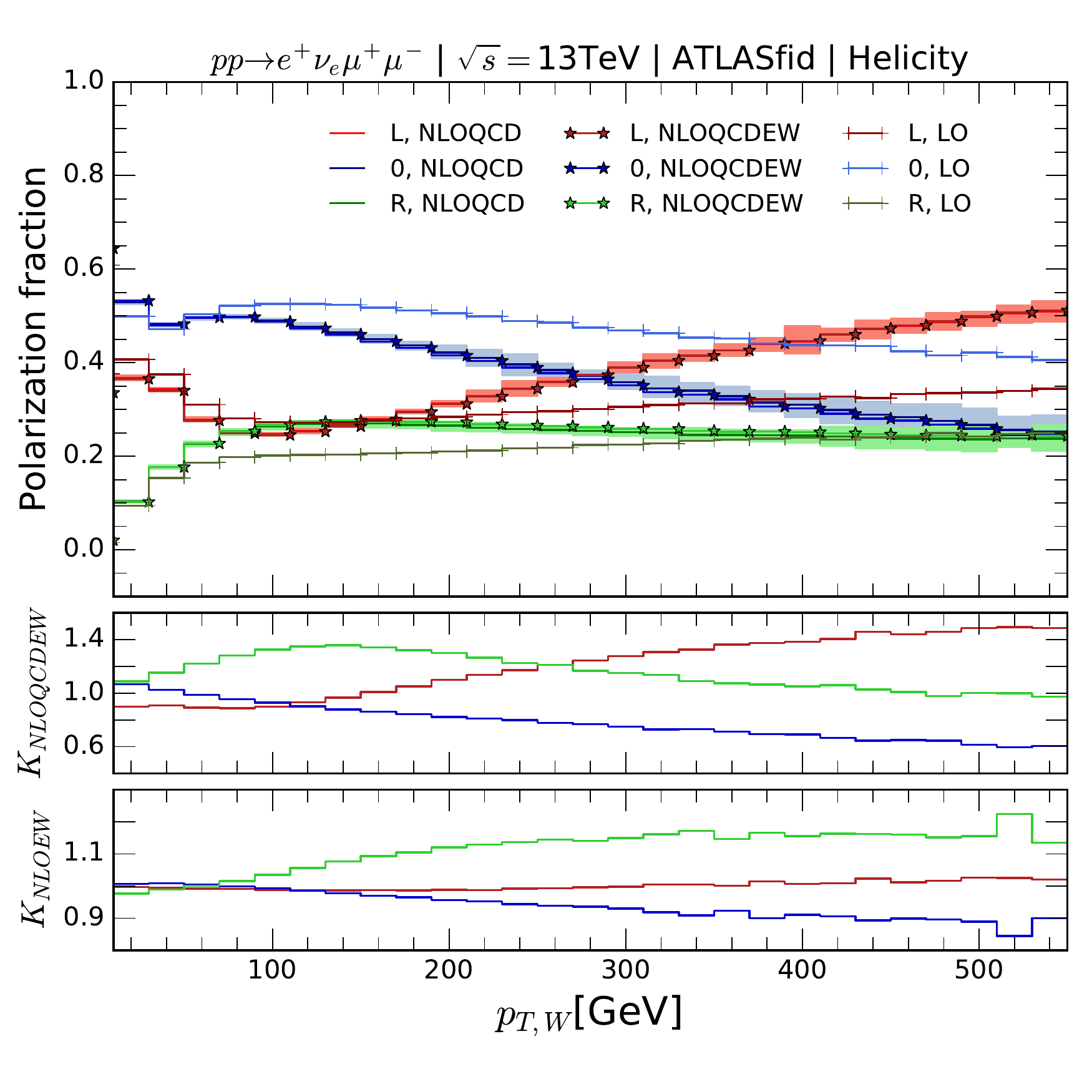}& 
    \includegraphics[width=0.48\textwidth]{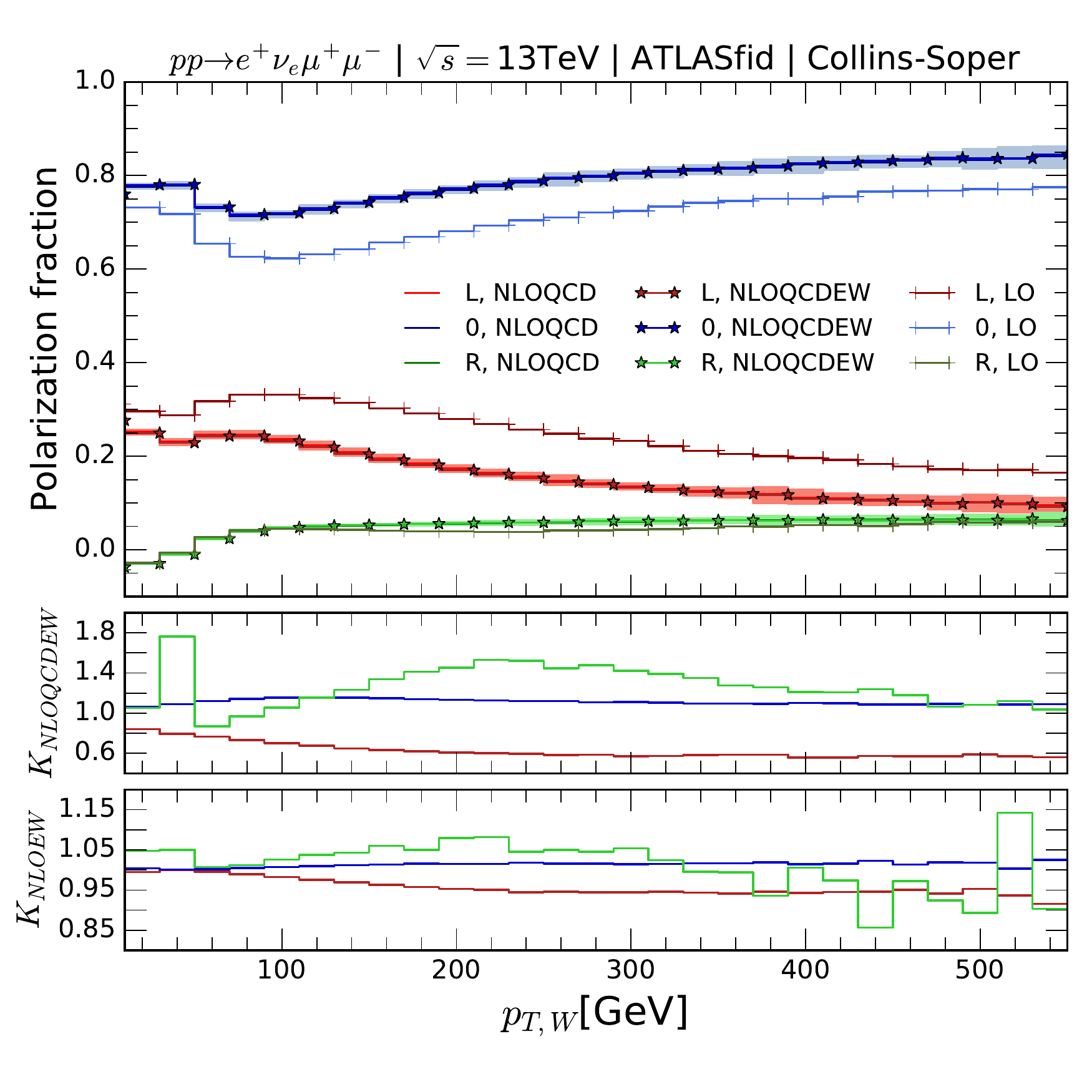}
  \end{tabular}
  \caption{Transverse momentum distributions of the $W^+$ boson
    fiducial polarisation fractions. The left-hand-side plot is for
    the helicity coordinate system, while the right-hand-side plot is
    for the Collins-Soper coordinate system. The bands include PDF and
    scale uncertainties calculated at NLO QCD using a linear
    summation. The $K$ factors in the small panels are the ratios of
    the NLO results over the LO ones.}
  \label{fig:dist_ptWp_ATLAS_NLOQCDEW}
\end{figure}

We also display in \fig{fig:dist_ptWp_ATLAS_NLOQCDEW} the $p_{T,W}$
distribution of the $W^+$ fiducial polarisation fractions calculated
from \eq{eq:polfrac}, in the HE (left) and CS (right) coordinate
systems. The two distributions have a very different behaviour: in the
CS system the fraction $f_R$ is negative at low $p_{T,W}$ and the
longitudinal fraction $f_L$ does not decrease at high $p_{T,W}$, while
all fractions are positive in the HE coordinate system and the
longitudinal fraction does decrease at high transverse momentum, in
line with the equivalence theorem and closer to the behaviour of
inclusive polarisation fractions. The same remark can be made for the
$Z$ polarisation observables~\cite{Baglio:2018rcu}.

\section{Conclusion}

We have presented in this contribution our calculation of the $W^\pm$
and $Z$ bosons polarisation observables, inclusively and at the
differential level, in the process $p p\to e^\pm \nu_e\mu^+\mu^- +X$
at the 13 TeV LHC including full NLO QCD corrections and NLO EW corrections
calculated in the double-pole approximation. This approximation has
been found to be in a very good agreement with the calculation of the
full EW corrections, while being much faster and easier to
calculate.

We have proposed the study of fiducial polarisation
observables that can be directly measured without using template fit
methods and give information about the underlying spin structure of
the $W$ and $Z$ bosons. The helicity coordinate system seems to be
more suitable than the Collins-Soper coordinate system as the
polarisation fractions are positive and decrease at high energies. EW
corrections are found to be important especially for the coefficients
$A_3^Z$ and $A_4^Z$ that are related to the $Z$ boson radiative
decay. We highly encourage the ATLAS and CMS experiment to perform the
measurement of these fiducial observables.

\section*{Acknowledgments}

JB acknowledges the support from the
Carl-Zeiss foundation and Moriond 2019 committee. This research is funded by the Vietnam
National Foundation for Science and Technology Development (NAFOSTED)
under grant number 103.01-2017.78. JB would like to thank the
organisers of the 54th Rencontres de Moriond EW session for the
invitation and the very fruitful atmosphere.

\section*{References}

% \begin{thebibliography}{99}

% \end{thebibliography}

%%%%%%%%%%  Bibliography  %%%%%%%%%%%%
\bibliographystyle{nb}
\bibliography{baglio_polarisation}

\end{document}